\documentclass{article}
\title{\bf 
1D Potts, Yang-Lee Edges and Chaos. 
}
\author{ {\it Brian P. Dolan}\\
Department of Mathematical Physics,\\
National University of Ireland,\\
Maynooth, Ireland\\        
\\
{\bf and}\\
\\
{\it D.A. Johnston}\\
         Dept. of Mathematics\\
         Heriot-Watt University\\
         Riccarton\\
         Edinburgh, EH14 4AS, Scotland
         }
\begin{document}
  \maketitle
                      {\Large
                      \begin{abstract}
%
It is known that the (exact)
renormalization transformations for the one-dimensional Ising model
in field can be cast in the form of a logistic map $f(x) = \lambda x ( 1 -x)$
with $\lambda=4$ and $x$ a function
of the Ising couplings $K$ and $h$. Remarkably, the line bounding the region of chaotic behaviour in $x$ is precisely that 
defining the Yang-Lee edge singularity 
in the Ising model. The generalisation of this
relation between the edge singularity and chaotic behaviour to other models
is an open question. 

In this paper we show that the one dimensional $q$-state Potts model for 
$q \ge 1$
also displays such behaviour. A suitable combination of couplings (which reduces to the Ising case for $q=2$) can again be used to define an $x$ satisfying
$f(x) = 4 x ( 1 -x)$. The Yang-Lee zeroes no longer lie on the unit circle
in the complex $z = exp ( h)$ plane for $q \ne 2$, but their locus is
still reproduced by the boundary of the chaotic region in the logistic map.
%
                        \end{abstract} }
%
  \thispagestyle{empty}
%
%
  \newpage
%
                  \pagenumbering{arabic}

\section{Introduction, Ising}

Yang and Lee \cite{YL1, YL2}, later followed by
various other authors \cite{others}, provided an important paradigm
for understanding the nature of phase transitions by looking at
the behaviour of spin models in {\it complex} external fields
They observed that the partition function of a system above its critical temperature $T_c$
was non-zero throughout some neighbourhood
of the real axis in the complex external field plane. As $T \rightarrow T_{c}+$
the endpoints of loci of zeroes moved in to pinch the real axis, signalling the transition.
When such endpoints occur at non-physical (i.e. complex) external field values they can
be considered as ordinary critical points with an associated edge critical
exponent. This appealing picture was later extended by Fisher to
temperature driven transitions 
\cite{fish}.

On any finite graph $G_n$ with $n$ vertices the free energy of an Ising-like spin model
can be written as
\begin{equation}
F(G_n,\beta,z) = - n h - \ln \prod_{k=1}^n ( z - z_k (\beta))
\end{equation}  
where the fugacity $z = \exp ( h)$, and $h$ is the (possibly complex) external field.
The $z_k (\beta)$ are the Yang-Lee zeroes, which in the thermodynamic limit
generally condense on curves in the complex $z$ plane. In the infinite
volume limit $n \rightarrow \infty$
the free energy per spin is
\begin{equation}
F(G_{\infty},\beta,z) = - h - \int_{-\pi}^{\pi} d \theta \rho(\beta , \theta) \ln ( z - e^{ i \theta} )
\end{equation}
where $\rho(\beta , \theta)$ is the density of the zeroes, which 
can be shown to appear on the unit circle
in the complex $z$ plane in the Ising case
(the Yang-Lee circle theorem).
For $T>T_c$ or, if one prefers $\beta<\beta_c$, there is a gap with
$\rho(\beta , \theta) = 0$ for $| \theta | < \theta_0$, and at these
edge singularities we have
\begin{equation}
\rho(\beta , \theta) \sim ( \theta - \theta_0 )^{\sigma}
\end{equation}
which defines the Yang-Lee edge exponent $\sigma$. This also
implies $M \sim ( \theta - \theta_0 )^{\sigma}$.
Various finite size scaling relations relate the Yang-Lee
exponent to the other critical exponents \cite{IPZ}
and can be used in numerical determinations of
critical behaviour \cite{enzo}. 

At first sight there is no apparent reason why the Yang-Lee edge singularity
should bear any relation to the onset of
chaotic behaviour in non-linear maps.
The relation between the two in the case of the 1D Ising model
was exposed in \cite{0} by considering the renormalization group
flow as an example of just such a mapping. The partition function for the 1D
Ising model
is given by
\begin{equation}
Z_N (K, h) =\sum_{\{\sigma\}} exp \left[ {K \sum_{j=1}^N \sigma_j \sigma_{j+1} + h
\sum_{j=1}^N \sigma_j} \right]
\end{equation}
where $K = {{J} \over {kT}}$ and $h = {{H} \over {kT}}$, with J the spin 
coupling
and H the external magnetic field, and periodic boundary conditions require
$
\sigma_{N+1} \equiv \sigma_1$. The well-known solution to the 1D Ising model
proceeds by expressing $Z_N(K,h)$ in terms of the transfer matrix $V$
as $Z_N = TrV^N$, where
\begin{equation}
V ( K, h) =\pmatrix{
V_{++}&V_{+-}\cr
V_{-+}&V_{--}\cr}
\quad = \quad
\pmatrix{
e^{K+h}&e^{-K}\cr
e^{-K}&e^{K-h}\cr
}
\label{eT}
\end{equation}

Diagonalising $V$ gives the eigenvalues
$
\lambda_\pm=e^K\left\{\cosh h\pm\sqrt{\sinh^2 h+e^{-4K}}\right\}
$ 
and allows us to express the partition function as
\begin{equation}
Z_N=\lambda_+^N + \lambda_-^N.
\end{equation}

The Yang-Lee zeroes of this partition function in the complex $h$ plane
are the $N$ roots of $Z_N(K,h)=0$, which are
for real $K$
the solutions of
\begin{equation}
Z_N = (\lambda_+)^N +(\lambda_-)^N = 0 \qquad\Leftrightarrow \qquad
\lambda_+ = \exp( {i n\pi\over N} )
\lambda_-
\end{equation}
where $-N < n \le N$ is odd. This gives
the $N$ Yang-Lee zeroes $h_n = i \theta_n$, 
\begin{equation}
\cos\Bigl({n\pi\over 2N}\Bigr)\sqrt{e^{-4K} + \sinh^2(h_n)}
=i\sin\Bigl({n\pi\over 2N}\Bigr)\cosh(h_n).
\label{YLedge0}
\end{equation}
which may also be rewritten as
\begin{equation}
\cos (\theta_n) = \sqrt{ 1 -e^{-4 K} } \cos \left( { n \pi \over 2 N} \right)
\label{YLedge}
\end{equation}
In particular we can see that when $K \rightarrow \infty$ (the zero
temperature ``transition point'' for the 1D Ising model) the zeroes are uniformly distributed on the unit circle in the complex $z$ plane, as demanded
by the Yang-Lee theorem.

So far, so standard. Now note that the recursive renormalization group transformation for the 1D Ising model can be obtained by demanding that any renormalised
couplings $K^\prime$ and $h^\prime$ satisfy
\begin{equation}
Z_{N \over 2} ( K^\prime, h^\prime ) = A^N Z_N ( K, h )
\end{equation}
where A is some renormalization factor. Thinking in terms of a decimation type
renormalization scheme it is clear that we can satisfy this by
taking 
\begin{equation}
V(K^\prime, h^\prime) = A^2 V( K, h)^2, 
\end{equation}
where $V$ is the transfer matrix given in
equ.(\ref{eT}). Viewed geometrically, we are welding two line segments
together and demanding a suitable rescaling of the couplings, so one
copy of the rescaled transfer matrix $V(K^\prime, h^\prime)$ must serve
in place of two copies of the original $V( K, h)$.
This leads to the recursion relations
\begin{eqnarray}
e^{2h^\prime} &=& e^{2h}\quad {\cosh (2K+h)\over\cosh(2K-h)} \nonumber \\
e^{4K^\prime} &=& \quad {\cosh(4K)+\cosh(2h)\over 2\cosh^2(h)}.
\label{RG}
\end{eqnarray}
The crucial observation of \cite{0} was that these recursion relations could
be recast by making use of the renormalization invariant
$m = 1 + e^{4 K} \sinh^2 (h)$ to eliminate $h$  and introducing the variable
\begin{equation}
x = - { m \over ( e^{4 K } - 1) }
\end{equation}
to  transform equs.(\ref{RG}) into the logistic map
$x^\prime = 4 x ( 1 - x)$. This will exhibit chaotic behaviour for $0 < x <1$,
i.e. if $m = 1 + e^{4 K} \sinh^2 (h) < 0$ which for imaginary external field,
$h = i \theta$, will occur if $\sin^2( \theta ) > e^{- 4 K}$.

What has this got to do with Yang-Lee edge singularities? Looking back at equ.(\ref{YLedge}) we can see that the lowest Yang-Lee zero will lie at
$\sin^2 ( \theta_0) = e^{- 4 K}$, which is precisely the ``boundary of chaos'',
$m<0$,
in $x$ observed in the renormalization transformation above. One can also 
identify a gap exponent for the chaotic map which is identical to the Yang-Lee
exponent $\sigma = -1 / 2$ for the 1D Ising model \cite{0}. It is natural to ask
whether the identification of the onset of chaos 
in an RG transformation and the Yang-Lee edge singularity is a peculiarity
of the 1D Ising model, or whether other examples of the phenomenon exist.
In the remainder of the paper we go to answer the question in the affirmative
for the 1D Potts model where one can also obtain the Yang-Lee zeroes explicitly
and construct an exact renormalization transformation along similar lines
to the Ising model.

\section{1D Potts, Yang-Lee}

The partition function for the 1D Potts model is given by 
\begin{equation}
Z_N (y, z) =\sum_{\{\sigma\}} exp \left[ {\tilde K \sum_{j=1}^N \delta( \sigma_j,  \sigma_{j+1}) + \tilde h
\sum_{j=1}^N \delta( \sigma_j, 1) } \right]
\label{ZPotts}
\end{equation}
where the $\delta()$s are Kronecker deltas and there are now $q$ possible
states for each spin $\sigma$. 
and we have defined $y = e^{\tilde K}$ and $z = e^{\tilde h}$ for 
later convenience.
We can write down a transfer matrix for this as a $q \times q$ matrix $V(y,z)$
with $q-2$ diagonal elements $( y - 1) / ( y z )^{1/q}$ and a $2 \times 2$
sub-matrix $T(y,z)$ \cite{Glue} \footnote{We have chosen a slightly
different form of the matrix $T(y,z)$ than \cite{Glue} for convenience
in formulating our renormalization group transformations. This simply
corresponds to different definitions of the ground state energy.} 
\begin{equation}
T ( y, z) = {1 \over ( y z )^{1/q}  }\pmatrix{
 y z & z^{1/2} ( q - 1) \cr
 z^{1/2} & y + q - 2 \cr}.
\label{eP}
\end{equation}   
For $q=2$ we recover $V(K,h)$ from $T(y,z)$
providing we identify  $ \tilde K  = 
2 K, \;  \tilde h =  2 h$.

The solution proceeds as in the Ising case by writing $Z_N (y,z) = tr V(y,z)^N$
and diagonalising $V$ \cite{Glue}. The dominant eigenvalues
$\lambda_{0,1}$ come from $T(y,z)$
\begin{equation}
\lambda_{0,1} = {1 \over 2} \left( ( y ( 1 +z ) + q - 2)   
\pm \sqrt{ (y ( 1 - z) + q - 2)^2 + ( q- 1) 4 z } \right) ( y z ) ^{ - {1 \over q}}
\end{equation}
which can be rewritten as
\begin{equation}
\lambda_{0,1} = {y \over 2} \left( t_+ t_- + z \pm \sqrt{ ( z - t_+^2 ) ( z - t_-^2)} \right) ( y z ) ^{ - {1 \over q}}
\end{equation}
with
\begin{equation}
t_{\pm} = {1 \over y} \left( \sqrt{ ( y -1 ) ( y + q -1)} \pm \sqrt{1 - q} \right).
\end{equation}
The other $q-2$ eigenvalues 
given by $\lambda_2 = \lambda_3 = \ldots =  ( y -1 ) ( y z )^{- { 1 \over q}}$
play no role in the thermodynamic limit.

The Yang-Lee zeroes $z_n = e^{h_n}$ , just as for the Ising model, appear as solutions of
\begin{equation}
Z_N = (\lambda_1)^N +(\lambda_0)^N = 0 \qquad\Leftrightarrow \qquad
\lambda_1 = \exp( {in\pi\over N} )
\lambda_0    
\end{equation}
which, upon substituting in the values above for $\lambda_{0,1}$, gives
\begin{equation}
\cos \left( {n \pi \over 2 N} \right) \sqrt{  ( z_n - t_+^2 ) ( z_n - t_
-^2)} = i \sin \left( {n \pi \over 2 N} \right) ( t_+ t_- + z_n )
\end{equation}
which is clearly of the same form as the Ising result in equ.(\ref{YLedge0})
for general $q$ and reproduces it exactly when $q=2$ (and we set 
$ \tilde K  =
2 K, \;  \tilde h =  2 h$), as it should. The resemblance runs deeper 
even for general $q$, as noted in \cite{Glue}. If we define 
$ \tilde z = z / ( t_+ t_- ) = y z / ( y + q - 2)$ this may be rewritten as
\begin{equation}
\cos \left( {n \pi \over 2 N} \right) \sqrt{  \left( \tilde z_n - { t_+ \over t_-} \right) \left( \tilde z_n - {t_- \over t_+ }\right)} = i \sin \left( {n \pi \over 2 N} \right) ( 1  + \tilde z_n )
\end{equation}
so in the complex $\tilde z$ plane the Yang-Lee zeroes are again uniformly
distributed round the unit circle as $\tilde K \rightarrow \infty$ and
$t_+ \rightarrow 1, t_- \rightarrow 1$.

We now try and pursue the same path with the renormalization group transformation
in the Potts model as for the Ising model.
We once more demand that renormalised
couplings $y^\prime$ and $z^\prime$ satisfy
\begin{equation}
Z_{N \over 2} ( y^\prime, z^\prime ) = A^N Z_N ( y, z )
\end{equation}
and again attempt to solve this by
taking $V(y^\prime, z^\prime) = A^2 V( y ,z )^2$, 
where $V$ is now the Potts transfer matrix. Since only $\lambda_{0,1}$
are playing any role in the thermodynamic limit we discard the remaining
$q-2$ eigenvalues and concentrate our attentions on the sub-matrix $T$,
by demanding $T(y^\prime, z^\prime) = A^2 T( y ,z )^2$. It is always
possible that an infelicitous choice of 
renormalization transformation could take us outside
the space of couplings spanned by $T$, but we shall see that this is
not the case here, at least for the symmetric choice of transfer matrix that
we have made.

We find the following recursion relations
\begin{eqnarray}
{1 \over  
 ( y^\prime z^\prime )^{1 \over q}} ( y^\prime z^\prime)
&=& {A^2 \over ( y z )^{2 \over q}} ( y^2 z^2 + z ( q-1)  )  \nonumber \\
 {1 \over 
 ( y^\prime z^\prime )^{1 \over q}} (y^\prime + q -2)  &=& {A^2 \over ( y z )^{2 \over q}}( z ( q - 1) + ( y + q - 2)^2 )\nonumber \\
{1 \over  ( y^\prime z^\prime )^{1 \over q}} (z^\prime)^{1/2}
&=& {A^2 z^{1/2} \over ( y z )^{2 \over q}}( z y +  y + q - 2 ) 
\end{eqnarray}
which can be used to eliminate $A$ giving
\begin{eqnarray}
\label{recursiontwo}
{y^\prime z^\prime \over y^\prime +q-2}&=&{y^2z^2+z(q-1) \over (y+q-2)^2+z(q-1)}  \nonumber \\ 
{(z^\prime)^{1\over 2} \over y^\prime +q-2}
&=&{z^{1/2}(yz + y+q-2) \over (y+q-2)^2+z(q-1)}.
\end{eqnarray}

It is then straightforward to show that, as for the Ising model, an invariant
exists --- in this case 
\begin{equation}
C= { (y ( 1 - z) + q - 2)^2 \over z} 
\label{Pottsinv}
\end{equation}
--- so we can use y and C to reduce our recurrence relations to one for $y$
alone.
Eliminating $z^\prime$ from (\ref{recursiontwo}) leads to a single
recursion relation which can be written as
\begin{equation}
\label{recursionone}
y^\prime(y^\prime + q -2) -(q-1) = 
{z\bigl( y(y +q-2) - (q-1)\bigr)^2 \over \bigl(y(z+1) +q-2\bigr)^2}.
\end{equation}
Now we can use (\ref{Pottsinv}) to write
\begin{equation} 
C + 4y(y+q-2)={\bigl( y(z+1) +q-2\bigr)^2 \over z}.
\end{equation}
So we define 
\begin{equation}
x=-{ [(C/4)+q-1] \over [(y-1)(y+q-1)]} 
\end{equation}
and the relation (\ref{recursionone}) is
again reduced to the logistic map with the pre-factor 4,
\begin{equation}
x^\prime = 4 x ( 1 - x).
\label{logistic}
\end{equation}
For C real and positive $x$ is real and negative and so is outside the
domain of chaos, but for $C< -4(q-1)$ $x$ is positive we have chaos
for $0<x<1$. On the critical line itself, $C=-4(q-1)$, we allow
ourselves the possibility of complex $z = | z | e^{i \theta}$ and find 
from equ.(\ref{Pottsinv}) that
\begin{equation}
z =  {(y+q-2) \over y} e^{i\theta} =  {( t_+ t_- )e^{i\theta}}
\label{zeq}
\end{equation}
where 
\begin{equation}
\cos (\theta )  = 1 - 2 {( q -1 ) \over y ( y + q - 2)}.
\label{coseq}
\end{equation}
Given our earlier discussion it is no surprise to find that these are
precisely the equations defining the Yang-Lee edge singularity
in the 1D Potts model. 

We have thus seen that defining a decimation type renormalization transformation
for the 1D Potts model gives rise to a set of recursion relations
which may be reduced using the renormalization invariant of equ.(\ref{Pottsinv})
to a single equation. This may in turn be mapped on to the logistic equation.
The boundary of the chaotic region for this logistic 
map is identical to the critical
line of the Yang-Lee edge singularity.  This behaviour is entirely analogous
to that seen in the 1D Ising model in \cite{0}.

Note that for complex temperatures $y$, and so $x$, is complex
and defining $w=-4(x-{1\over 2})$ turns (\ref{logistic}) 
into the Mandelbrot map on the complex plane,
\begin{equation}
w^\prime = w^2 -2.
\end{equation}

\section{Discussion}

The similarity of the Yang-Lee edge singularity for the general $q$
state  Potts models
in 1D and for the Ising model was already remarked in \cite{Glue}.
The exponent $\sigma = - 1 /2$ is identical for all $q>1$, and
in suitably rescaled variables the Yang-Lee zeroes lie on the
unit circle at $T=0$ for all $q>1$.
Given this, it is
perhaps not so surprising that the relation between the 
renormalization transformations and the Yang-Lee edge singularity
also survive to general $q$ from the Ising model. 
Nonetheless, it is intriguing that an exactly soluble model has again demonstrated a close relation between the onset of chaos in a renormalization map
and the Yang-Lee edge singularity.

We have not discussed the case $0 \le q <1$
in this paper, since \cite{Glue} makes it clear that
the correspondence with the Ising model is rather less direct for this.
For $0 \le q<1$ the Yang-Lee edges $z_{+} , z_{-}$,
lie on the positive real $z$ axis for all temperatures. Inside
this interval  points exist where a third eigenvalue $\lambda_2$ comes
into play, which would involve the extension of our renormalization
transformation to a $3 \times 3$ matrix. For $q=1$ on the other hand, 
the discussion in the previous section suggests that the
critical line for the logistic map is 
defined by $C=0$ and equ.(\ref{Pottsinv}) then shows that this
translates to 
$z = (y -1) / y$ and $\theta=0$ 
which defines a circle
of radius $( y - 1) / y$ in the $z$ plane with a ``gap angle'' $\theta=0$. 
Direct consideration in \cite{Glue}
of equs.~(\ref{zeq},\ref{coseq})
gives the same result for the Yang-Lee
zeroes. This demonstrates that the boundary of chaos
in the renormalization map
and the Yang-Lee edge singularity are one and the same for $q=1$ also.

In a wider context pathologies of 
approximate real space renormalization transformations
{\it within} phases 
in higher dimensional models
have been discussed in a rigorous manner in \cite{enter}.
Other cases exist where recursive non-linear maps are used 
in the definition of exact partition
functions, notably for spin models on Bethe lattices (trees). There has been
discussion of chaotic effects in such models when $q<1$ \cite{Monroe}
and the logistic equation has even been observed for a $q=1$ state Potts
model (related to percolation) on a Bethe lattice with co-ordination number
3 \cite{Wagner}. Similarly, there have been extensive investigations
of the effects of frustration in inducing chaotic behaviour 
in such maps for spin models on trees \cite{A3,A5,A5,A6,M2} but 
no discussion of any Yang-Lee/chaos link in this context.
Since results are available for the Yang-Lee edge singularity
in the Ising model
on $\phi^3$ random graphs \cite{desedge} 
\footnote{These are equivalent to the Bethe lattice 
as far as the (mean-field) spin 
model critical behaviour is concerned \cite{desloops}},
it would be worth investigating the correspondence, if any,
between chaos and the Yang-Lee edge singularity in this case also.
This would make it clearer whether the phenomenon reported in
\cite{0} and here was merely a 1D quirk, or something more general.

As a final remark we note that the 1D Potts models possess a 
temperature/field (i.e. $y,z$) duality \cite{Glue}
\begin{eqnarray}
y^D &=& { z + q -1 \over z - 1} \nonumber \\
z^D &=& { y + q -1 \over y - 1}
\end{eqnarray}
so the relation between the chaotic behaviour discussed here and the endpoints
of Yang-Lee (field) zeroes can also be couched in terms of Fisher (temperature)
zeroes in the dual $y^D, z^D$ variables. 

\section{Acknowledgements}

B.D. was partially supported by Enterprise Ireland Basic Research Grant - SC/1998/739 and B.D. and D.J. by an Enterprise Ireland/British 
Council Research Visits Scheme - BC/2000/004. 
B.D. would like to thank Heriot-Watt mathematics department for its 
hospitality and D.J. would like to thank
the Department of Mathematical Physics,
National University of Ireland,
Maynooth for the same.
 
\bigskip

\end{document}